\documentclass[sigconf]{acmart}

\usepackage{graphicx}
\usepackage{xurl}
\usepackage{csquotes}
\usepackage[inline]{enumitem}
\usepackage{array}
\usepackage{float}
\newcolumntype{P}[1]{>{\raggedright\arraybackslash}p{#1}}

\AtBeginDocument{%
  \providecommand\BibTeX{{%
    \normalfont B\kern-0.5em{\scshape i\kern-0.25em b}\kern-0.8em\TeX}}}

\copyrightyear{2023} 
\acmYear{2023} 
\setcopyright{rightsretained} 
\acmConference[FAccT '23]{2023 ACM Conference on Fairness, Accountability, and Transparency}{June 12--15, 2023}{Chicago, IL, USA}
\acmBooktitle{2023 ACM Conference on Fairness, Accountability, and Transparency (FAccT '23), June 12--15, 2023, Chicago, IL, USA}
\acmDOI{10.1145/3593013.3593994}
\acmISBN{979-8-4007-0192-4/23/06}

\begin{document}

\title[Certification Labels for Trustworthy AI]{Certification Labels for Trustworthy AI: Insights From an Empirical Mixed-Method Study}

\author{Nicolas Scharowski}
\email{nicolas.scharowski@unibas.ch}
\orcid{0001-5983-346X}
\affiliation{%
  \institution{University of Basel}
  \streetaddress{}
  \city{}
  \state{}
  \country{}
  \postcode{}
}

\author{Michaela Benk}
\email{mbenk@ethz.ch}
\orcid{0002-8171-320X}
\affiliation{%
  \institution{Mobiliar Lab for Analytics \\ ETH Zürich }
  \streetaddress{}
  \city{}
  \country{}}

\author{Swen J. Kühne}
\email{swen.kuehne@zhaw.ch}
\orcid{0001-8326-9168}
\affiliation{
  \institution{School of Applied Psychology \\ Zürich University of Applied Sciences}
  \streetaddress{}
  \city{}
  \state{}
  \country{}
  \postcode{}}

\author{Léane Wettstein}
\email{leane.wettstein@unibas.ch}
\orcid{0003-5068-7007}
\affiliation{
  \institution{University of Basel}
  \streetaddress{}
  \city{}
  \state{}
  \country{}
  \postcode{}}

\author{Florian Brühlmann}
\email{florian.bruehlmann@unibas.ch}
\orcid{0001-8945-3273}
\affiliation{
  \institution{University of Basel}
  \streetaddress{}
  \city{}
  \state{}
  \country{}
  \postcode{}}
  
\renewcommand{\shortauthors}{Scharowski et al.}

\begin{abstract}

Auditing plays a pivotal role in the development of trustworthy AI. However, current research primarily focuses on creating auditable AI documentation, which is intended for regulators and experts rather than end-users affected by AI decisions. How to communicate to members of the public that an AI has been audited and considered trustworthy remains an open challenge. This study empirically investigated \emph{certification labels} as a promising solution. Through interviews ($N = 12$) and a census-representative survey ($N = 302$), we investigated end-users' attitudes toward certification labels and their effectiveness in communicating trustworthiness in low- and high-stakes AI scenarios. Based on the survey results, we demonstrate that labels can significantly increase end-users' trust and willingness to use AI in both low- and high-stakes scenarios. However, end-users' preferences for certification labels and their effect on trust and willingness to use AI were more pronounced in high-stake scenarios. Qualitative content analysis of the interviews revealed opportunities and limitations of certification labels, as well as facilitators and inhibitors for the effective use of labels in the context of AI. For example, while certification labels can mitigate data-related concerns expressed by end-users (e.g., privacy and data protection), other concerns (e.g., model performance) are more challenging to address. Our study provides valuable insights and recommendations for designing and implementing certification labels as a promising constituent within the trustworthy AI ecosystem.

\end{abstract}

\ccsdesc[500]{Human-centered computing~Empirical studies in HCI}

\keywords{AI, Audit, Documentation, Label, Seal, Certification, Trust, Trustworthy, User study}

\maketitle

\section{Introduction}

In recent years, the promise of artificial intelligence (AI) in transforming our lives has seen widespread advances in all sectors of society. AI is increasingly guiding our consumer choices \cite{Puntoni2020ConsumersAA}, reshaping service by automatizing tasks \cite{Huang2018ArtificialII}, assisting managers in hiring decisions \cite{Li2021AlgorithmicHI}, or augmenting clinical decision-making \cite{Yu2018ArtificialII}. In light of increasingly ubiquitous AI and its profound impact on human lives, various government institutions, scientific communities, and the general public are engaged in a widespread discourse on how to ensure trustworthy AI \cite{jacovi2021formalizing, Liao-et-al_2022, Jobin2019TheGL, Kaur2022TrustworthyAI} for both low-, and high-stake scenarios \cite{whitepaperAI}.

To this end, a large body of work has focused on identifying the principles that underlie trustworthy AI \cite{Kaur2022TrustworthyAI}. They include mitigating bias and unfairness in AI systems \cite{Lepri2018FairTA}, explaining the reasoning of AI decisions \cite{langer2021we}, setting up mechanisms to hold AI accountable \cite{Kaur2022TrustworthyAI}, and ensuring user privacy \cite{stahl2018ethics}. However, as trust is determined by people's perception \cite{Liao-et-al_2022, lee2004trust}, efforts to design trustworthy AI are hampered by a lack of understanding of how to communicate trustworthiness to people, for instance, through documentation or other transparency affordances \cite{Liao-et-al_2022}. Particularly for end-users\footnote{In line with prior work \cite{langer2021we, wang_fairness, Seifert2019}, we define end-users in this paper as laypeople (i.e., non-experts in data science or machine learning) who may be affected directly or indirectly by the outcomes of AI systems.}, trusting AI can be a challenge, as they lack the necessary expertise and knowledge to evaluate the various trustworthiness principles (e.g., robustness, privacy, fairness) \cite{Bhatt2020ExplainableML, Knowles-et-al_2021}.

Motivated by these challenges, this work builds on research highlighting the pivotal role of \textit{auditability} as an enabler of trust in AI \cite{toreini2020relationship, Brundage2020} and its crucial role in creating an "AI trustworthiness ecosystem" \cite{avin2021filling} by ensuring that the principles of trustworthy AI are met. Auditing refers to mechanisms that evaluate and ensure compliance with regulations and ethical standards \cite{Raji2020}. Various methods have been proposed to increase AI systems' transparency and, thereby auditability, such as through the use of model documentation or information about datasets \cite{Crisan2022InteractiveMC, Gebru2021}. While AI documentations are valuable artifacts to inform audit decisions, they are tailored to regulators and experts and not intended to certify and communicate to end-users that an AI has met the auditing criteria.

For this reason, our work focuses on communicating the outcomes of auditing processes to end-users, a topic that has received little attention in previous work. Specifically, we investigate the use of \textit{certification labels}, which are commonly used in other domains, such as food and energy \cite{taufique2022revisiting, EULabel, Ecocert2023}. Certification labels are relevant in the context of trustworthy AI for three reasons. First, through the use of simple language, icons, or color-coding, they are usually designed to be accessible to various stakeholder groups, including end-users with limited knowledge and time \cite{grunert2014sustainability}. Second, if reflecting a genuine and credible auditing process, certification labels can communicate the criteria used in an audit, thereby serving as a "trustworthiness cue" for end-users \citep{liao2020questioning, schlicker2022calibrated}. Third, labels have shown to promote trustworthiness of a product in other domains \cite{tonkin2015trust} facing similar challenges on how to certify that a product meets certain criteria, such as agricultural standards (e.g., organic foods \cite{Ecocert2023}) or low ecological impact (e.g., sustainable hotels \cite{EULabel}). However, end-users' attitudes toward AI certification labels and their effectiveness in communicating the trustworthiness of AI remain to be explored. 

We addressed this gap by conducting a mixed-method study with both interviews ($N = 12$) and a census-representative survey ($N = 302$) with end-users. Our results provide evidence that certification labels can effectively communicate AI trustworthiness. Qualitative findings revealed that end-users have positive attitudes toward AI certification labels and that labels can increase perceived transparency and fairness and are regarded as an opportunity to establish standards for AI systems. Particularly, data-related concerns expressed by end-users, such as privacy and data protection, can be mitigated through the use of certification labels. However, labels may not be able to address all raised concerns, such as model performance, suggesting that they should be considered one promising constituent among others for trustworthy AI. Furthermore, our results provide insights into facilitators and inhibitors for the effective design of certification labels in the context of AI. For example, end-users expressed strong preferences for independent audits and highlighted the challenge of communicating subjective criteria such as "fairness," whose meaning can be ambiguous. 

Quantitative findings showed that a certification label significantly increases end-users' trust and willingness to use AI in both low- and high-stake AI scenarios. Nevertheless, end-users reported a higher preference for certification labels in high-stake scenarios (e.g., hiring procedure) than in low-stake scenarios (e.g., price comparison), and the positive effect of a label on trust and willingness to use AI was more pronounced in high-stake scenarios. This suggests that compliance with mandatory requirements for AI in high-stake scenarios could be effectively communicated to end-users through certification labels in addition to the proposed voluntary labeling for low-stake AI scenarios \cite{whitepaperAI, stuurman2022regulating}. 

To summarize, our study is the first to demonstrate the potential of certification labels as a promising approach for communicating to end-users that an audit has certified an AI to be trustworthy. We contribute to the trustworthy AI literature by highlighting opportunities and challenges for designing and effectively implementing certification labels.

\section{Auditing for trustworthy AI}

A growing body of work recognizes the critical role of algorithmic or AI auditing in enabling the trustworthiness of AI systems \citep{toreini2020relationship, avin2021filling, Knowles-et-al_2021}. Prior work suggests that auditing improves fairness \cite{Wilson_21}, accountability \cite{Constanza_22}, and governance \cite{falco2021governing}, among others. These elements are considered to contribute to trust in and acceptance of AI\footnote{The definition of trust in AI and its operationalization is an ongoing debate \cite{Vereschak2021HowTE, jacovi2021formalizing, toreini2020relationship, scharowski2022trust}. As an extensive theoretical discussion is out of scope of this work, we focus on trustworthiness, a property of the trustee, rather than on trust as a process that can be affected by numerous contextual and personal factors \cite{castelfranchi2010trust, Chiou2021TrustingAD}.}. Moreover, audits have the ability to expose problematic behavior, such as algorithmic discrimination, distortion, exploitation, and misjudgment \cite{Bandy_21}. 
In safety-critical industries such as aerospace, medicine, and finance, audits are a long-standing practice \cite{Constanza_22}. However, only recently have researchers recognized that these areas could inform AI auditing and acknowledged the importance of considering insights from the social sciences, where audits have emerged from efforts toward racial equity and social justice \cite{Vecchione_21}. 

While the importance of AI auditing has been identified, the development of common audit practices, standards, or regulatory guidance is ongoing \cite{Bandy_21, Constanza_22} and efforts to create auditing frameworks throughout the AI development life-cycle are still in their early stages \cite{Raji2020}. Auditing can be defined as "an independent evaluation of conformance of software products and processes to applicable regulations, standards, guidelines, plans, specifications, and procedures." \cite[p. 30]{audit}. At least three types of AI auditing can be distinguished, including first-party internal auditing, second-party audits conducted by contractors, and independent third-party audits \cite{Constanza_22}. However, whether auditing should be conducted by independent third-parties or internally within organizations is a topic of ongoing academic discussion \cite{falco2021governing, Raji2020, Krafft_21}, with both approaches having their advantages and drawbacks. \citeauthor{Raji2020} argue that external auditing may be constrained by a lack of access to organizations' internal processes and information that are often subject to trade secrets. In contrast, \citeauthor{falco2021governing} point out that the outcomes of internal audits are typically not publicly disclosed and that it often remains unclear whether the auditor's recommendations are effectively implemented or not. The question of whether end-users prefer internal or external audits remains to be investigated.

In addition to defining standards and best practices for AI auditing, it is crucial to consider how the outcomes of audits can be communicated to different stakeholders with varying knowledge and needs \cite{Yurrita_22}. Current research has mainly focused on approaches for documenting machine learning (ML) models and training datasets. These artifacts play an important role in the AI trustworthiness ecosystem by increasing transparency and allowing auditors and regulators to determine whether principles of trustworthy AI (e.g., fairness, robustness, privacy \cite{Kaur2022TrustworthyAI}) have been met \cite{Knowles-et-al_2021}. For example, "model cards" \cite{Mitchell-et-al_2019, Crisan2022InteractiveMC} disclose information about a model's purpose and design process, its underlying assumptions, and the model's performance characteristics. Similarly, \citeauthor{Gebru2021} introduced "datasheets," which summarize the motivation, composition, collection process, and recommended uses for datasets, and \citeauthor{floridi2022capai} recommended the use of "summary datasheets" and "external scorecards." The former is aligned with the goals of "datasheets" and synthesizes key information about the AI, including its purpose, status, and contact information. The latter is conceptually closely related to "model cards" and evaluates the AI system along several dimensions to form an overall risk score \citep{floridi2022capai}. 

However, these documentations are tailored to AI practitioners, and regulators \cite{Yurrita_22, Seifert2019, Knowles-et-al_2021}, rather than end-users affected by AI decisions. Often, end-users have neither the access nor the expertise to understand the technical information that AI documentation provides \cite{arnold2019factsheets}. It is unlikely that end-users can effectively utilize ML model documentation or data documentation to make informed judgments about trusting or using AI \cite{Knowles-et-al_2021}. For this reason, end-users depend on auditors and regulators who can use these artifacts to verify and ensure the trustworthiness of AI. Yet, it remains an open research question of how to effectively communicate to end-users that an audit has considered an AI trustworthy. End-users require accessible communication tailored to their specific values and concerns \cite{Yurrita_22}. A potentially effective way to provide such information is through the use of \emph{certification labels}, which we will introduce in the following.  

\section{Certification labels for audited AI}

Labels are widely used for displaying specific product or service attributes to help consumers make more informed decisions. They are well-established in various fields, such as agriculture \cite{gorton2021determines}, food \cite{jones2019front}, energy \cite{stadelmann2018different}, and e-commerce \cite{thompson2019trustmarks}. Different kinds of labels exist, and various classification systems have been proposed \cite{stuurman2022regulating, taufique2022revisiting, ikonen2020consumer}. For example, in the food industry, "nutrition labels" provide consumers with simplified and easily understandable information to identify a product's nutritional content. While this information can also be found in detailed tables on the back of food packing, for many consumers, this information is too complex, revealing similar challenges end-users face with AI documentation. This is where labels can provide information in a clear and accessible manner, utilizing simple language, icons, and color coding, which makes labels accessible to individuals from different backgrounds \cite{goodman2018impact, grunert2014sustainability}. Prior work in consumer research has shown that labels can communicate the outcomes of audits and thereby enhance trust in a product \cite{tonkin2015trust}.

In this study, we focus on \emph{certification labels}, which certify that a product or service meets one or several criteria and are thus suitable for the case of audited AI. Certification labels are exclusively awarded to products that have undergone an auditing process, typically conducted by a third-party organization \cite{taufique2022revisiting}. By communicating an institutional assurance of trustworthiness, third-party organizations can serve as "trust surrogates" for the consumer, shifting the trust relation from trust in the AI to trust in the institution that provides the certification \cite{tonkin2015trust}. In this case, a certification label serves as a trustworthiness cue \cite{schlicker2022calibrated} that signals compliance with governance structures. Our work thus closely aligns with the proposal by \citeauthor{Liao-et-al_2022}, highlighting that the trustworthiness of AI is not inherently given but must be communicated and perceived as such by the user, for instance, through transparency affordances. According to the authors, people then use heuristics (i.e., mental rules of thumb) to evaluate these affordance cues to form judgments about the trustworthiness of AI. The authors further suggest that certifications from regulatory bodies that have audited the AI could serve as trustworthiness cues, invoking these heuristics. Therefore, certification labels in the context of AI are a promising approach to communicate that a regulatory body has audited an AI and considered it trustworthy. 

There have been several initiatives at a national and international level to introduce AI labels in both industry (e.g., \cite{forhumanity}, \cite{aieig}, \cite{fraunhofer}) and government (e.g., \cite{danishLabel}, \cite{maltaLabel}). These initiatives vary in their intended scope but are mostly still in an early stage. Previous studies have also emphasized the potential of labels as a means of AI certification \cite{stuurman2022regulating, Seifert2019, Holland_nutrition20}. \citeauthor{Holland_nutrition20} proposed the concept of a "Data Set Nutrition Label," which would summarize key aspects of a dataset (e.g., metadata and the data source) prior to the development of ML models. \citeauthor{Seifert2019} further suggested labels for trained ML models that independent reviewers have evaluated based on properties such as accuracy, fairness, and transparency. A recent study by \citeauthor{stuurman2022regulating} commented on various labels to provide information to end-users affected by AI decisions. Drawing from the EU Act on AI \cite{euAIAct}, the study distinguished between low-stake and high-stake AI systems and proposed a voluntary labeling system for AI not considered high-stake. This distinction aligns with recommendations from the EU's "white paper on artificial intelligence," \cite{whitepaperAI} which encourages organizations to use labels to demonstrate the trustworthiness of their AI-based products and services. A survey conducted with individuals and organizations directly or indirectly engaged in audits found that while respondents believed that AI audits should be mandatory, 53\% supported mandating them only for high-stakes systems \cite{Constanza_22}. End-users' perceptions of certification labels in low and high-stakes AI scenarios have not yet been investigated.

Despite this extensive theoretical work on labels in the context of AI and their gradual adoption in industry and government, there is currently a lack of empirical research exploring end-users' attitudes toward AI certification labels and their effectiveness in communicating trustworthiness in low- and high-stake AI scenarios. This study aims to address this research gap and inform current industry and government initiatives.

\section{Research Questions}

Based on the aforementioned considerations, we investigated the following research questions:

\begin{itemize}
    \item[\textbf{RQ1:}] What are end-users' attitudes toward certification labels in the context of AI? 
    \item[\textbf{RQ2:}] How do certification labels affect end-users' trust and willingness to use AI in low- and high-stake scenarios?
\end{itemize}

\section{Methods}

To answer these research questions, we used a mixed-method research approach consisting of semi-structured interviews and a subsequent survey to collect quantitative data as part of a within-subjects design study. For both the interviews and the survey, we used a scenario-based approach to investigate people's attitudes and the effects of a certification label, inspired by past research \citep{kapania2022because, Jakesch, Binns.2018}. In the interviews, we asked participants about their attitudes toward AI and certification labels. As a follow-up within-subjects study, we implemented a survey to investigate the effect of a certification label quantitatively. The semi-structured interviews served as a basis for the survey and a means to enrich the quantitative results. The quantitative survey complemented the qualitative interviews by extending our results to a larger census-representative sample. In the following, we will introduce the certification label used in our study before describing the procedures of each method in more detail.

\subsection{The certification label}

To investigate labels in the context of AI, we used a certification label that has already been developed for the broader context of digital trust. Using an existing label had the advantage that it had undergone an extensive design process and thus did not need to be created from scratch.

\vspace{-8mm}
\begin{figure} [!htbp]
    \includegraphics[width=\linewidth]{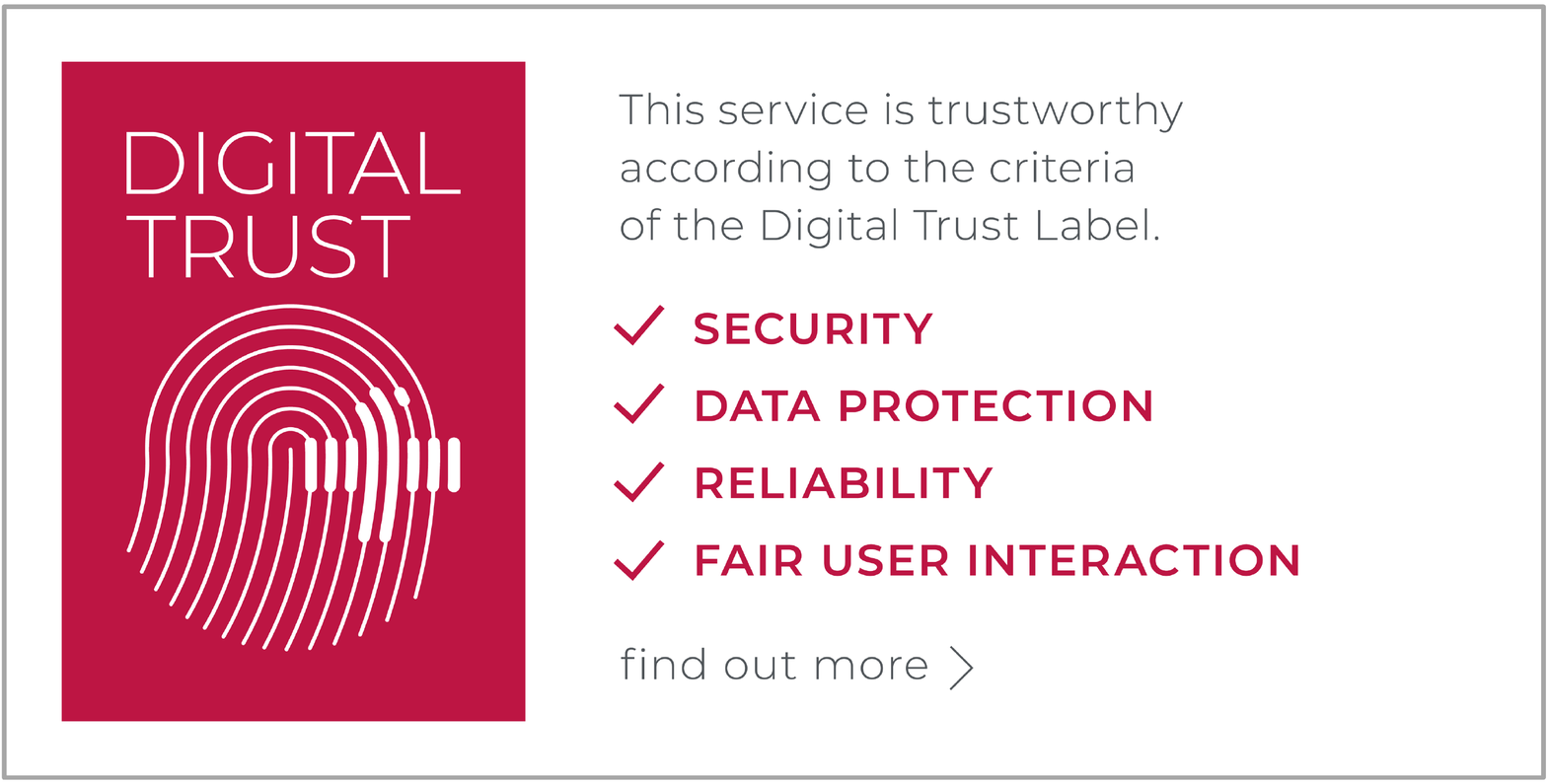}
    \vspace{-15mm}
    \caption{The "Digital Trust Label," which we adopted as a certification label for AI. ©2023 Swiss Digital Initiative}
    \label{fig:sdi_label}
    \Description{X} 
\end{figure}

The non-profit foundation Swiss Digital Initiative laid the groundwork for developing this certification label. At the label's core lies a catalog of verifiable and auditable criteria, co-developed by an academic expert group based on a user study on digital trust. A panel of independent experts from academia, data and consumer protection, and digital ethics further developed the label catalog. Involving digital service providers and auditors in the designing process ensured that the criteria were auditable and verifiable. The catalog that forms the basis of the audit currently contains 35 criteria that are summarized into four categories:

\begin{enumerate}
    \item Security (criteria 1 - 12): What is the security standard? The service provider shall, e.g., ensure that the data is encrypted as it transfers so that third-parties cannot access it.
    \item Data protection (criteria 13 - 20): How is the data protected? The service provider shall, e.g., assume responsibility for the appropriate management of the data.
    \item Reliability (criteria 21 - 29): How reliable is the service or product? The service provider shall, e.g., take all actions required to safeguard the continuity of the service. 
    \item Fair user interaction (criteria 30 - 35): Is automated decision-making involved? The service provider shall, e.g., ensure that all users receive equal treatment and that there is no data-based service or price discrimination.
\end{enumerate}

If an organization would like its digital product or service (e.g., a chatbot) to receive the certification label, it can voluntarily request an audit and thus participate in the certification process. After a scoping call with third-party auditors, an audit is performed along the criteria catalog. The audit leads to an audit report detailing the performance per criterion, which is double-checked by an independent label certification committee composed of auditing experts. If non-conformities are identified, the organization applying for the label must fix the identified issues, e.g., adjust its privacy policy. After a successful auditing report, the certification label is awarded for a period of three years with two audits during that period.

\subsection{Scenario selection}

Participants were presented with real-world examples of AI systems, adapted from \citeauthor{kapania2022because}, namely \textit{medical diagnosis, loan approval, hiring procedure, music preference, route planning} and \textit{price comparison} (see materials on OSF: \url{https://osf.io/gzp5k/}). One advantage of using hypothetical scenarios instead of real consumer applications is that differences in participants' prior experience with the applications can be controlled for \citeauthor{kapania2022because} and \citeauthor{woods2006comparing} proposed that people's behavior in scenario-based experiments corresponds to their real-life behavior. To answer our second research question and following \citeauthor{kapania2022because} we explored both low-stake scenarios (music preference, route planning, price comparison) and high-stake scenarios (medical diagnosis, hiring procedure, loan approval). This distinction was crucial since other researchers \cite{floridi2022capai, stuurman2022regulating} and the "EU AI Act" \cite{euAIAct} have discussed the use of AI labels for "low-stake" and "high-stake" scenarios. This classification was based on the AI's respective impact on affected parties and the involvement of significant risks, in particular with respect to safety, consumer rights, and the use of personal data.

\subsection{Interviews}

\subsubsection{Participants}

Initially, we invited 16 participants to an interview on-site at the university. The recruitment was carried out through a university-internal database and an online marketplace where scientific studies can be advertised. 
To ensure that our sample consisted of end-users (i.e., laypeople who may be affected directly or indirectly by the outcomes of AI systems), we used screening questions following \citeauthor{kapania2022because} and asked potential participants about their knowledge of AI and experience working with AI-based systems. We selected participants who indicated that they have heard about AI but did not work with it and provided a comprehensible description or adequate example of what AI is without overly restricting the valid responses (e.g., "robots" was valid while obvious nonsense answers such as "E.T. the alien" was deemed invalid). In addition, we asked participants to indicate their age, gender, profession, and English language proficiency so that we could design the interviews as balanced as possible and present materials in English.
However, four interviews did not take place due to no-shows. We, therefore, conducted 12 interviews with end-users of different backgrounds, ages, and genders that lasted 60 - 90 minutes. The interviews were conducted in German and recorded through field notes and audio recordings. Each participant received compensation in the form of a gift card worth CHF 10.00 from a Swiss retail company. 
The final sample ($M_{age} = 35.42$, $SD_{age} = 12.50$, $Min_{age} = 23$, $Max_{age} = 66$) consisted of students (P2, P3, P4, P8, P11) enrolled in linguistics and literature (P2), fine arts (P3), and psychology (P4, P8, P11), as well as individuals who described their occupation as a bike messenger (P12), waitress (P1), dancer (P9), course manager (P7), management assistant (P6), intern (P10) and retired teacher (P5). The sample was predominantly female, with ten women and two men.

\subsubsection{Procedure}

Before the interviews, participants had to read and sign a declaration of consent. In the declaration, we informed participants of the purpose and rationale of the study, the researcher affiliations, the voluntary nature of study participation, and how their data will be analyzed and shared. All personally identifiable information was deleted to ensure privacy, and the anonymous data was stored without actual reference to the participants.

During the interviews, we asked attitudinal questions about AI, specifically where participants saw opportunities and challenges in using AI. We then presented the six scenarios to the participants without specifying the low- and high-stake categorization we had made in advance. Based on the respective headings of the scenarios (e.g., music preference), without further information, we asked participants to order the scenarios via drag and drop from "most impactful" (rank 1) to "least impactful" (rank 6). To ensure comparability, we defined "most impactful" for participants as "the scenario that would have the greatest impact on your personal life." This question aimed to validate our categorization in low- and high-stake scenarios. Next, we presented participants with one low-stake and one high-stake scenario and asked how they differed from one another. After this, participants were introduced to the certification label and asked how they perceived it, whether the label criteria were comprehensible or not, and where they saw opportunities and drawbacks of a certification label. The goal of the interviews was not only to gather qualitative data, but also to identify and determine which questions best suited the subsequent survey. We, therefore, made sure the questions were comprehensible and free of ambiguities. Any difficulties encountered during the interviews were discussed within the research team, and, if necessary, the respective questions were revised or removed. We refer to the digital repository for the complete interview manual.

\subsection{Survey}

\subsubsection{Participants}

To gain insights into how a general population perceives a label in the context of AI, we hired a market research agency (\url{https://www.bilendi.ch/}) to provide us with a Swiss census-representative sample regarding age and gender (quota sampling). We used the same screening questions as in the interviews and initially recruited 395 participants that received CHF 3.00 for taking part in the 15-minute online survey. Following a quality assessment using a self-reported single item as an indicator of careless responding \citep{meade2012identifying, bruehlmann2020quality}, 302 participants remained for data analysis. The sample is census-representative regarding age ($M_{age} =  43.88$, $SD_{age} = 16.08$, $Min_{age} = 18$, $Max_{age} = 79$) and the gender distribution (150 women, 151 men, one non-binary person).

\subsubsection{Procedure and measures}

The survey consisted of three parts.
First, after providing informed consent and a brief introduction to the study, participants were free to select one scenario from the low-stake and one from the high-stake categorization. After making their choice, they received full descriptions of the two scenarios (see \autoref{appendix:a}) and were asked to rate their trust ("how much would you trust the AI in the scenario presented?") and willingness to use ("how much would you be willing to use the AI in the scenario presented?") on a scale from 0 (= not at all) to 100 (= absolutely). In addition, participants were asked in which scenario they would more readily accept the AI's decision/recommendation (i.e., "in which of the two scenarios would you be more willing to accept the decision/recommendation made by AI?").

Participants were introduced to the certification label in the second part of the survey. They were asked for their impression and rated the importance of each criterion (i.e., "how important are the label criteria for you in the context of AI?") on a scale from 0 (= not at all) to 100 (= absolutely). Participants were also asked what effect the certification label had on their acceptance (i.e., "would you be more likely to accept an AI's decision/recommendation if it had received a label?") and preference (i.e., "in which one of the two scenarios would you prefer the use of a label?"). To understand end-users' preferences regarding external and internal auditing, we included an open-ended question (i.e., "who do you think should be responsible for awarding such a label?").

Finally, in the fourth part, we again let participants rate the AI in the same low- and high-stake scenario on trust and willingness to use, this time with the information that the AI had been awarded a certification label. This second assessment allowed us to examine the certification label's effect on trust and willingness to use ratings. Similarly to the first assessment, we asked participants to justify their ratings and why a label led to increased/decreased or unchanged ratings. At the end of the survey, we asked the participants for feedback and the question, "\textit{in your honest opinion, should we use your data in our analyses in this study? Do not worry, this will not affect your payment. You will receive the compensation either way}," as an additional quality check. The complete survey can be found on the digital repository.

\subsection{Analysis and coding procedure}

We used the qualitative interview data to answer RQ1 and the quantitative survey data to answer RQ2. The interview data was evaluated using qualitative content analysis \citep{mayring2019qualitative}, more specifically summarizing content analysis. We followed the procedure according to \citeauthor{mayring2019qualitative} by determining the coding unit, paraphrasing, generalization to the level of abstraction, first reduction, and second reduction to form a cross-case category system. Coding was carried out by three researchers who independently went through four interviews each. To ensure consistency, one interview was evaluated by all researchers. Any ambiguities and discrepancies were resolved through open discussions, and the final cross-case category system was formed in a group session. The quantitative data analysis was carried out in R (version 4.2.2. \cite{rcore}). We used the \textit{ggstatsplot} package (version 0.9.1. \citep{patil2021statsexpressions}) to conduct statistical testing and report $t$-values, standard deviations, and the corresponding $p$-values. We set the level of statistical significance to $\alpha = .05$.

\section{Results}

\begin{table*}[htbp]
\footnotesize
\renewcommand{\arraystretch}{1.25}
\centering
\caption{End-users' attitudes toward certification labels}
\label{tab:my-table}
\begin{tabular}{ P{2.5cm} P{2.5cm} p{9cm} } 
\textbf{Category} &
  \textbf{Subcategory} &
  \textbf{Example quote} \\ \hline
Opportunities for certification labels &
  Increasing trust &
  \textit{"Because if it is monitored and these various criteria have to be met in order to get the label, then I as a consumer can, of course, trust better and also know that there are perhaps controls and random checks, so I would definitely trust more." (P6)} \\
 &
  Increasing perceived transparency &
  \textit{"I think that if there is such an established label, it will certainly help to increase transparency." (P6) } \\
 &
  Increasing perceived fairness &
  \textit{"With the Fair User Interaction aspect, yes, probably so [fairness is increased]. … if the AI is now checked for this, and it can be determined that it is not data-based, treated differently." (P12)} \\
 &
  Auditing of AI systems &
  \textit{"Because I'm not an expert in the field and the label …, gives me proof … that it's tested by experts." (P4) }\\
 &
 Establishing standards for AI systems &
  \textit{"So I could imagine that if it is a bit more standardized, so to speak, because you have to meet certain standards, that it could introduce a general level of fairness." (P3)}
\\ 
  &
  Covering relevant concerns &
  \textit{"The concern [responsibility] was covered and then just the general concern with all just how our data is also used and hopefully not misused, or yes. That is also covered." (P10)} 
  \\
  \hline
  Facilitators for effective certification labels &
  Additional label information & 
  \textit{"[I would like to] find out what this "Fair User Interaction" means, what it refers to, how my data is protected … how is it designed and who monitors this label. Exactly by whom was it created and by whom it is administered, awarded and so on, that's what I would like to know." (P12)}
   \\
 &
  Independent party awarding the label & 
  \textit{"Ideally, it would be an overarching body that is, for example, also external and has the competences and the knowledge … ideally, an NGO that runs it without any vested interest." (P12)}
   \\
 &
  Recognition of label &
  \textit{"If many companies get involved in using this label. Then I think it could have an impact." (P9)}
  \\
  &
  Clarity of label criteria &
  \textit{"[The criteria] are totally comprehensible to me, in any case. It's also something that would be important to me if I were to use such a program." (P9) }
  \\
  &
  Actuality of label &
  \textit{"You could say that the label guarantees that work on AI is ongoing." (P11)}
   \\ 
   \hline
Limitations of certification labels &
  Unaddressed concerns & 
  \textit{"What you could include is a criterion for the AI. That an AI has been used enough times and has, for example, been 99\% correct and always had the right answers, rather than 80\%." (P4) }
   \\
 &
  Lack of persuasiveness & 
  \textit{"I think there are still a lot of people, or some people, who will be critical of these systems even though it has a label." (P3)}
   \\
   \hline

Inhibitors for effective certification labels &
 Overabundance of labels & 
 \textit{"Because you can see that in the organic sector, there are now 20 labels and as a consumer you can almost no longer categorize them, so I think it's so important now that there is also Bio-Suisse [an organic label] or something like that in Switzerland, they have established themselves well, but I think you always have to stick to that as a label." (P6) }
   \\
 &
  Vacuousness of label criteria &
  \textit{"I find these four points are so common. And bad news is, maybe we don't really analyze what is written. Or don't even read. I can't speak of everyone, but speaking of myself. I often just don't read that message. Beautiful words, but all blah blah blah." (P2)}
   \\ 
   &
   Subjectivity of label criteria &
   \textit{"Yes, so what is complete transparency? That brings us back to fairness … what is fair? These are all such subjective terms that, in my eyes, you can't use like in natural sciences - where you calculate and then there's a result - it's soft science where you're working in." (P5) 
   }
   \\
   &
   Overlaps of label criteria &
   \textit{"Overlap; I think it all goes a bit in a similar direction, except maybe the last point [Fair User Interaction], which is a bit different again." (P10)}
   \\ \hline
\end{tabular}
\end{table*}

\subsection{Attitudes toward certification labels}

The content analysis of the interview data resulted in 127 case-specific categories, which were further consolidated across participants into 25 categories. These cross-categories were grouped into the following topics: \emph{"AI-related concerns, risks, problems,"}, \emph{"AI-related opportunities, advantages,"}, \emph{"attitudes toward certification labels,"}, and perceived \emph{"differences between low- and high-stakes scenarios"}. For the purpose of this study, we focus on the topic "attitudes toward certification labels," as this was the most relevant to our current research objective. Categories may consist of further subcategories. \autoref{tab:my-table} contains the subcategories and corresponding example quotes from end-users' attitudes toward certification labels. The complete content analysis with all topics is available on the digital repository.

\subsubsection{Opportunities and facilitators}

Participants in the interview study indicated that the label covered essential concerns. The content analysis revealed that the topic "concerns, risks, and problems" predominantly consisted of data-related concerns such as data privacy (i.e., protecting data from attack and malicious use), data storage (i.e., how data is handled and stored), and third-party involvement (i.e., unwanted and unknown disclosure of data). Regarding data-related concerns, a certification label for AI systems was perceived as an effective tool to convey compliance with these requirements and hold the certified parties more accountable. In particular, the security and data protection criteria were perceived as minimal standards that must be met for them to consider using AI. Participants emphasized that a certification label provides a certain level of transparency that removes the burden of examining these criteria from end-users. In addition, they viewed the certification labels and corresponding auditing process as an opportunity for more fairness and to establish standards for AI systems, allowing them to compare products and services critically. The interviewed participants indicated that a certification label could increase their trust for all these reasons. 

For a label to be convincing, participants emphasized that additional information regarding the label is needed. This includes information about the label's criteria (i.e., how were they formed?), the auditing process itself (i.e., how were these criteria weighted?), and the auditors (i.e., who was responsible for awarding a label?). Participants also placed a strong emphasis on the independence of the auditing process, noting that the auditors should have no financial ties to or other direct dependencies on the organizations for whose products or services the label is awarded in order not to undermine their credibility. Additionally, participants stressed the importance of widespread participation in the auditing and certification process, as this was deemed necessary for adopting AI standards and the label's credibility. As a crucial factor for the effectiveness of a certification label, participants identified regular updates that align with industry standards and best practices to ensure that the label remains relevant and useful.

\subsubsection{Limitations and inhibitors}

While participants acknowledged that a certification label covers essential issues, they also noted that it does not address all their AI-related concerns. These concerns included the lack of model performance (e.g., accuracy measures). Some participants noted that a certification label alone could even lead to "blind trust" in AI systems without accuracy measures. Additionally, participants noted that while a certification label provides some level of transparency, it does not provide complete documentation (e.g., source code) of the AI system and the ethical reasoning behind the auditors' decision to approve the use of AI in a particular application in the first place. As a result of these limitations, participants felt that a certification label might not be sufficiently persuasive to convey trustworthiness for critical individuals. 

Furthermore, participants identified several reasons why a certification label may not be effective. One reason was a potential overabundance of labels with different standards, diluting compliance with regulations and leading to confusion among end-users. In line with this, participants emphasized the importance of ensuring that the label's criteria are not just "empty promises" but that they are actually adhered to by organizations. They also pointed out the difficulty of measuring the label's criteria and the degree of subjectivity involved. Concepts such as security and fairness can mean different things to different people. Results showed that some criteria were more easily understood (e.g., security) than others (e.g., fair user interaction). For example, 11/12 participants implied that the definition of the security criteria covered what they had in mind. For data protection, this was the case for 9/12 participants, followed by 8/12 participants for reliability. However, merely 2/12 participants indicated that the criterion "fair user interaction" captured what they thought it would encompass. In addition to these differences in comprehension, participants pointed out conceptual overlaps for some criteria (e.g., security and data protection) that were not readily understood without further clarification. All these factors might diminish the effectiveness of a certification label.

\subsection{Effects of certification labels}

\begin{figure*} [!htbp]
    \includegraphics[width=13.2cm]{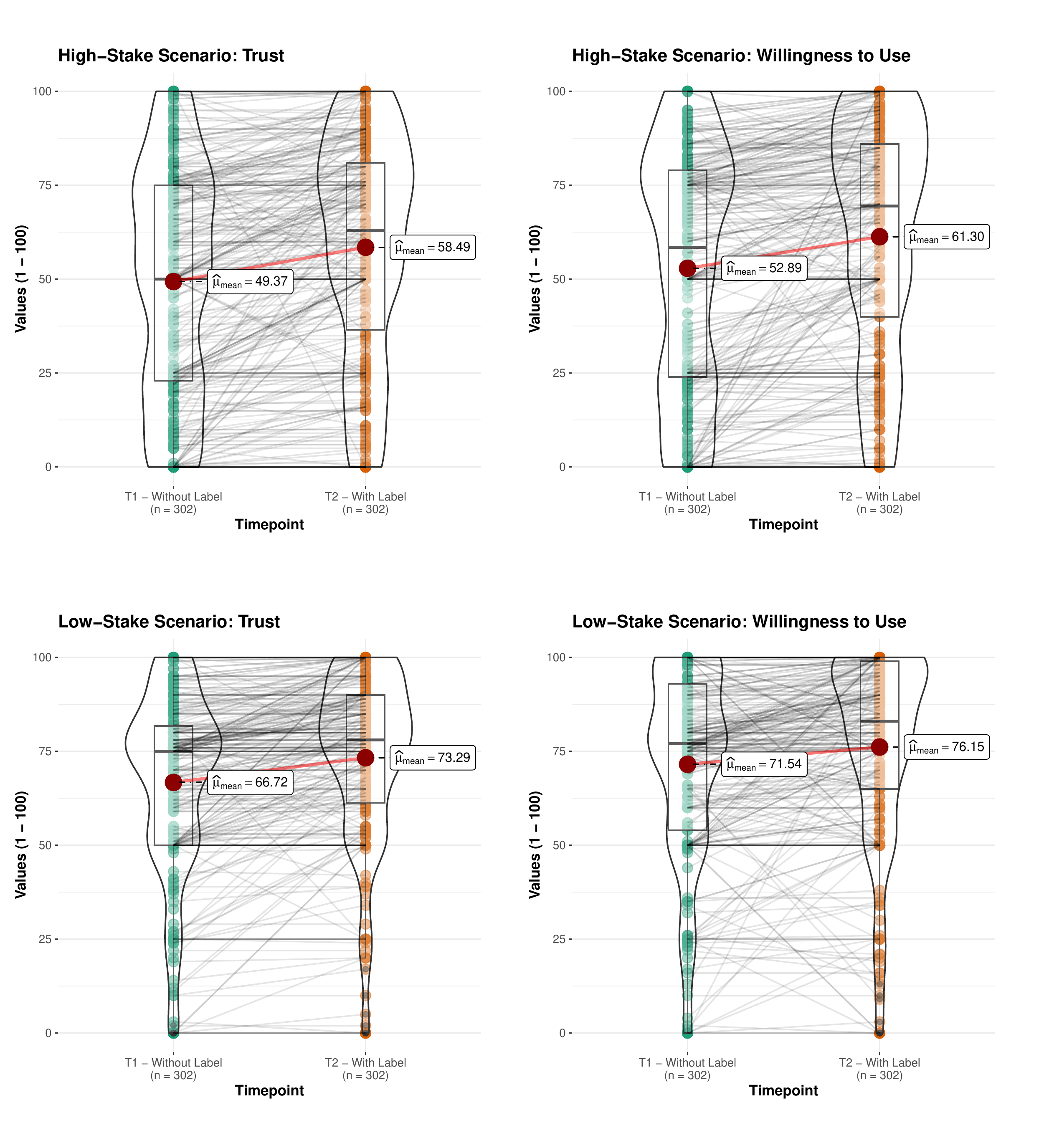}
    \vspace{-5mm}
    \caption{Plots showing the individual scores for trust and willingness to use and their respective changes from T1 (without label) to T2 (with label). The plots also depict the medians, means, and distribution of the aggregated low- and high-stake scenarios. All comparisons revealed statistically significant differences.}
    \label{fig:trust_use_change}
    \Description{X} 
\end{figure*}

Participants in the survey study were asked to select one case each from the high-stake (medical diagnosis, hiring procedure, loan approval) and one from the low-stake (music preference, route planning, price comparison) scenarios without explicitly being informed of this distinction. Validation of this distinction between low- and high-stake was provided by participants' "impactfulness" rankings. Calculating a mode revealed that the three high-stake scenarios were perceived as the most impactful ones (i.e., 1 = medical diagnosis, 2 = hiring process, 3 = loan approval, 4 = price comparison, 5 = music preference, 6 = route planning). The majority of participants indicated that they would be more likely to accept the AI's decision/recommendation in low-risk scenarios (74.2\%, $n = 224$) than in high-risk scenarios (17.9\%, $n = 54$) and 7.9\% ($n = 24$) indicating no preference, which we considered an additional confirmation of the distinctiveness of the two scenarios. Participants in the interview study distinguished between low- and high-stakes scenarios primarily on the level of risk associated with the scenario. They reported that high-stakes scenarios carry higher self-relevance and long-term consequences. 

Before being presented with the certification label, participants reported both higher trust ($M = 66.72$, $SD = 24.27$) and willingness to use ($M = 71.54$, $SD = 25.54$) ratings for the low-stake scenarios, compared to ratings in high-stake scenarios for trust ($M = 49.37$, $SD = 30.76$) and willingness to use ($M = 52.89$, $SD = 32.63$). 
After being presented with the certification label, participants' trust and willingness to use ratings revealed statistically significant increases in both low- and high-stakes scenarios (see \autoref{fig:trust_use_change}). 
A dependent Student's $t$-test indicated that the presence of a certification label resulted in the highest increase for trust ($M_\Delta = 9.12$, $SD = 17.92$, $t(301) = 8.84$, $p < .001$) and willingness to use ($M_\Delta = 8.41$, $SD = 17.69$, $t(301) = 8.26$, $p < .001$) ratings in high-stake scenarios, followed by trust ($M_\Delta = 6.57$, $SD = 13.26$, $t(301) = 8.61$, $p < .001$) and willingness to use ($M_\Delta = 4.60$, $SD = 17.03$, $t(301) = 4.70$, $p < .001$) ratings in low-stake scenarios. 
Hedges' $g$ for effect sizes ranged between .27 - .51 and can thus be considered small (for low-stake scenarios) to medium (for high-stake scenarios) \cite{sawilowsky2009new}.

\begin{figure*} [htbp]
    \includegraphics[width=13.5cm]{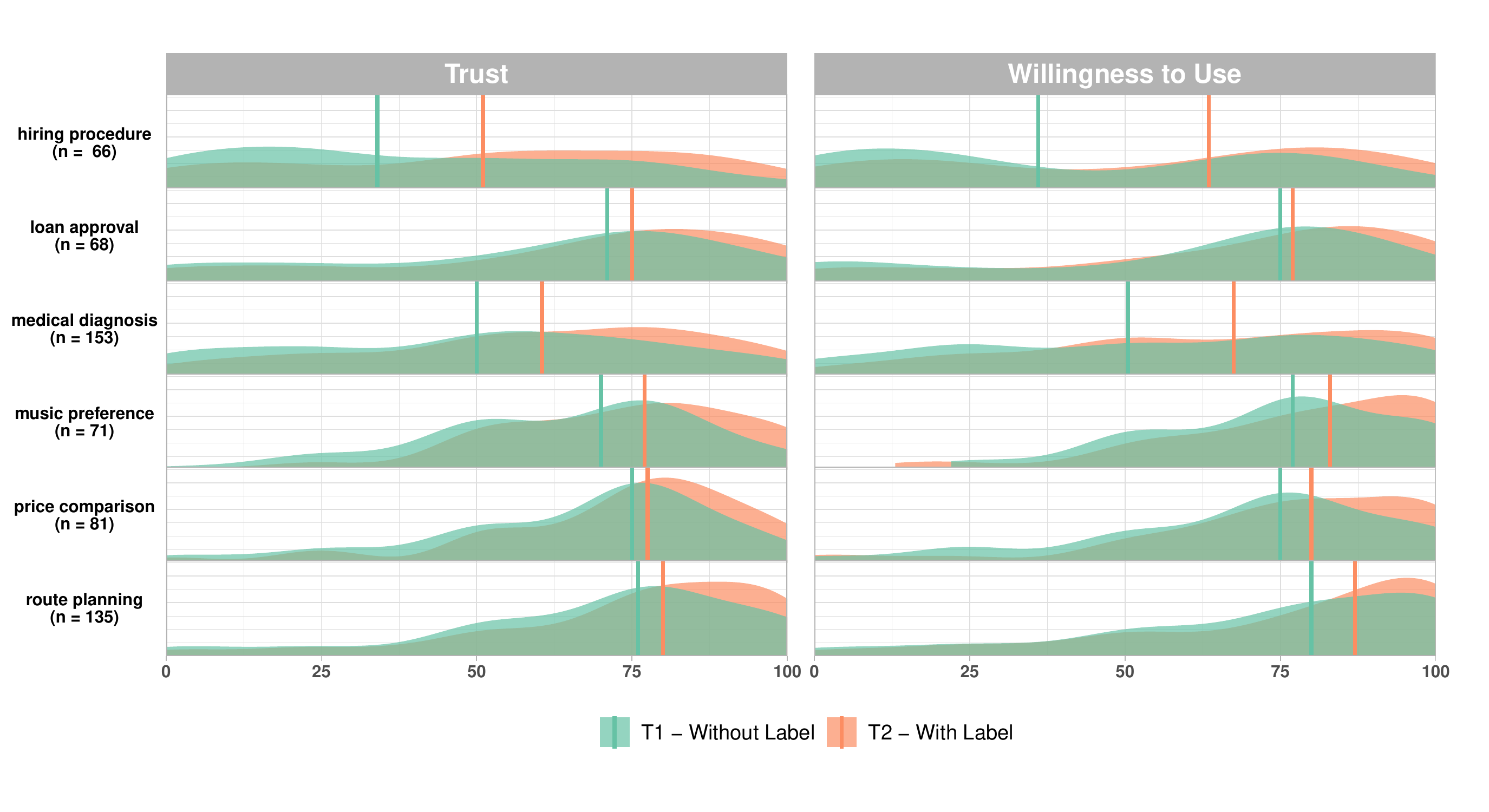}
    \vspace{-5mm}
    \caption{Plots showing the different distributions for trust and willingness to use ratings for the different high-stake (hiring procedure, loan approval, medical diagnosis) and low-stake (music preference, price comparison, route planning) without a label at T1 and with a label at T2.}
    \label{fig:scenario_distribution}
    \Description{X} 
\end{figure*}

The different ratings depending on low- and high-stake scenarios become evident when considering the violin plots and boxplots (see \autoref{fig:trust_use_change}). The ratings for high-stake scenarios are relatively symmetrically distributed across the scale. In contrast, the low-stake scenarios' distribution is heavily left-skewed, with approximately 75\% of the data above a rating of 50 for trust and willingness to use. Introducing a certification label for both scenarios leads to a further shift of the distribution to the right and, thus, higher ratings. 
Plotting the non-aggregated scenarios individually reveals the distributional differences more clearly (see \autoref{fig:scenario_distribution}). The ratings of the individual high-stakes scenarios are more spread out on the scale than in the case of the low-stake scenarios. Differences in the effectiveness of a label also become apparent from this perspective. The median trust and willingness to use ratings in all scenarios increases in the presence of a label and are more pronounced in the high-stake scenarios.

A majority of the survey participants directly indicated that they would prefer the use of a certification label in the selected high-stake scenario (63.2\%, $n = 191$), compared to preferring a label in the low-stake scenarios (22.2\%, $n = 67$), with 14.6\% ($n = 44$) of participants indicating no preference. Regarding the different preferences for certification labels in low- and high-stake scenarios, participants from the interview study expressed a greater demand for a certification label in high-stake scenarios because of the higher scenario complexity, limited individual expertise, and a lack of prior experience with the system. Overall, 81.1\% ($n = 245$) of survey participants stated a preference for using an AI with a label, compared to 6\% ($n = 18$) that would prefer to use an AI without a label and 12.9\% ($n = 39$) that stated no preference. Also, 70.9\% ($n = 214$) indicated to be more likely to accept an AI's decision/recommendation if it had received a label, compared to 14.2\% ($n = 43$) that indicated "no," and 14.9\% ($n = 45$), that made no statement. Survey participants rated the importance of the existing label criteria in the context of AI at a high level with similar ratings for security ($M = 87.72$, $SD = 20.93$), data protection ($M = 85.04$, $SD = 21.81$), reliability ($M = 76.97$, $SD = 23.19$) and fair user interaction ($M = 80.80$, $SD = 23.37$). However, merely 55.3\% ($n = 167$) of the participants agreed that the label addresses the concerns/challenges/risks they see that come with the use of AI, while 20.9\% ($n = 63$) stated "no" and 23.8\% ($n = 72$) indicated that no statement was possible. 

When being asked the question of who should be responsible for awarding a label, the open-ended responses from the survey revealed that a majority of participants expressed a preference for external entities to conduct the auditing, with 48.7\% ($n = 147$) of the answers being coded as "government" and 37.4\% ($n = 113$) as "NGO." Only 5.3\% ($n = 16$) of the answers were coded as "company." Additionally, 8.6\% ($n = 26$) of the responses were coded as "other," which included mentions of entities such as "ethic committee," "consumer protection," or "citizen's association."

\section{Discussion}

The quantitative findings reveal that the presence of a certification label significantly increases participants' trust and willingness to use AI in \emph{both} low- and high-stake scenarios, thereby answering our second research question. Most participants (81\%) of the census-representative survey preferred using AI with a certification label, and a large proportion of participants (71\%) responded that they would be more likely to accept an AI's decision or recommendation if it had been awarded a certification label. The results further show that a majority of participants (63\%) not only indicated a preference for certification labels in high-stake scenarios, but that certification labels also had a larger effect on trust and willingness to use AI in high-stake scenarios. For example, willingness to use ratings for the "hiring procedure" scenario increased from 36 to 64 points, compared to an increase from 75 to 80 points for the "price comparison" scenario. While \citeauthor{stuurman2022regulating} and the EU's "white paper on artificial intelligence" distinguish between regulating high-stake AI through mandatory requirements and proposed voluntary labeling only for low-stake AI, our results demonstrate the relevance of certification labels for end-users, specifically in high-stake scenarios. Based on these findings, we argue that parallel to voluntary labeling for low-stake AI scenarios, compliance with mandatory requirements for AI in high-stake scenarios could also be communicated through certification labels, potentially increasing end-users' trust in and willingness to use awarded AI systems.

Qualitative findings allowed us to answer our first research question and provide a more nuanced picture of which aspects to consider for effective certification labels in the context of AI. The certification label we investigated in this study was designed for digital trust more generally. However, end-users' attitudes toward the certification label were primarily positive, and the label's criteria of security, data protection, reliability, and fair user interaction were also relevant to end-users in the context of AI. We derive this from survey participants' high "importance" ratings for the existing label criteria. Concerning \emph{opportunities} for AI labels, participants in the interview study indicated that a certification label could increase perceived transparency and fairness and serve as a means to establish standards for AI systems. It became apparent from the interviews that certification labels can especially cover end-users' data-related concerns (e.g., privacy, data protection, and third-party involvement) that map to previous work \cite{toreini2020relationship}. 

However, our results also reveal that certification labels have \emph{limitations} and do not alleviate all issues end-users face regarding the use of AI. Only half of the participants in the survey indicated that a certification label addresses their AI-related concerns/challenges/risks, suggesting that end-users seem to hold differentiated needs. For example, participants in our interviews pointed out that a certification label does not provide indicators about the AI's performance (e.g., accuracy measures). They remarked that performance indicators are essential in deciding in which cases the AI can be trusted and when it must be questioned. This led participants to remark that a label could inadvertently foster "blind trust" if performance indicators are absent. Thus, we suggest that certification labels should either include performance indicators as part of the label criteria or be supplemented with them. Based on these results, we argue that certification labels can more readily signal trustworthiness than untrustworthiness. This is because it is not possible to distinguish if a digital product or service has not yet been audited or whether it has failed to meet specific audit criteria, particularly if certification labels remain voluntary. We regard certification labels as \emph{one} component of an "AI trustworthiness ecosystem" \cite{avin2021filling} that meets essential needs for end-users but which ideally should be combined with other transparency approaches to signal untrustworthiness (e.g., accuracy measures) and form a "chain of trust" \cite{toreini2020relationship}.

As potential \emph{inhibitors} for effective certification labels, participants in our interviews pointed out certain overlaps and the subjective nature of the label's criteria. Ultimately, "fairness" and "security" are subjective judgments that vary from one person to the next, and our results showed that the criterion "fair user interaction," in particular, did not reflect what study participants thought it encompassed. The challenge for auditing of defining and measuring concepts that are inherently difficult to quantify has been discussed by previous research \cite{Seifert2019, Vecchione_21, Knowles-et-al_2021}. Our results indicate that this subjectivity is recognized by end-users and can impair the effectiveness of a label. To avoid a discrepancy between, for example, the auditors' definition of fairness and what people commonly associate with this term, auditors should be in dialogue with end-users so that their values are represented in a label. This is in line with \citeauthor{Constanza_22}, who had criticized that the involvement of affected communities plays a minor role in AI audits. They argued that real-world harms and sociological phenomena could only be understood by engaging with people to inform auditing.

Our interview results highlight that end-users request not only information on the label's criteria but also information regarding the criteria content (i.e., how they were formed), the auditing process itself (i.e., how the criteria informed the audit), and particularly about the auditors (i.e., who awarded the label). We identified this demand for additional information as a potential \emph{facilitator}, indicating that an effective certification label is more than just a list of evaluation criteria. A large majority (86\%) of survey participants responded that either the government (49\%) or a non-governmental organization (37\%) should ideally be responsible for awarding a label, with only 5.3\% of responses indicating that a company should be responsible. Participants in the interview study emphasized the auditors' independence (e.g., financially, with no conflict of interest) as a prerequisite for the effectiveness of a certification label. These findings support the notion that auditing can only foster trust if the auditors themselves are trusted \cite{avin2021filling} and are in line with results of label studies in other domains \cite{gorton2021determines, tonkin2015trust}, which show that third-party certification positively affects trust in eco-labels. We contribute to the ongoing discussion regarding internal vs. external auditing by showing that end-users favor independent auditors. To account for this independence on the one hand and the structural advantages of internal audits on the other, "cooperative audits" \cite{Wilson_21} could be a way forward, balancing between the advantages and challenges of the two approaches. In addition to these facilitators and inhibitors, auditors and regulators should also be mindful that an overabundance of labels with different standards can inhibit the persuasiveness and trustworthiness of their certification label. Such effects have been reported for eco-labels, where an extensive number of existing labels result in different standards that remain unclear to consumers \cite{harbaugh2011label}. These findings speak for a certain harmonization and regulation of certification labels. Moreover, organizational compliance with a label's criteria should be established so end-users do not perceive them as "empty promises" but instead as a means for increased accountability for organizations and more trustworthy AI \cite{Knowles-et-al_2021}. A prominent instance of such a challenge is the case of the CE (conformité européenne) marking, in which some products use the mark without actually being manufactured to EU quality standards \cite{CE_Label}. This illegitimate use has led, among other things, to the introduction of supplementary certification labels to certify product quality, which unintentionally contribute to consumer confusion \cite{stuurman2022regulating}. To realize their full potential, certification labels should have a thorough auditing process, be regularly updated to reflect current industry standards, and ideally, be used by a wide range of organizations to increase recognition.

\section{Limitations and future work}

We conducted a within-subjects survey study where participants were presented with the AI scenarios with and without a certification label. While this provided valuable insights into the general effectiveness of certification labels, future work could compare label classes or designs (e.g., nutrition labels vs. certification labels) in a between-subjects experimental design. Certification labels are limited in their ability to communicate untrustworthiness. While other kinds of labels have a more differentiated rating system (e.g., color-codings or grades) that allows comparisons, certification labels only provide dichotomous information by either being present or not. Thus, it is not possible to differentiate if a product without a certification label is untrustworthy because it failed to meet a label's criteria or has yet to be audited. A between-subjects design could provide evidence about the effectiveness of different kinds of labels and identify the factors that make labels more or less effective in communicating trustworthiness and untrustworthiness.

Moreover, we used single-item questions to measure trust and willingness to use. Trust, in particular, is a complex psychological construct \cite{scharowski2022trust} and might not be adequately operationalized using single-items measures. However, a recent study has shown that single-item trust measures are equivalent to validated questionnaires regarding sensitivity to changes in trust and a reliable tool in longer surveys where questionnaires are not feasible \cite{nesset_22}. Future work should confirm the effectiveness of certification labels in fostering trust with validated psychometric measures and explore their effect on trusting dynamics that emerge over time in real-world human-AI interactions. 

\section{Conclusion}

This study empirically investigated certification labels to communicate trustworthy AI to end-users. For this purpose, we explored end-users' attitudes toward certification labels in the context of AI and how labels affect trust and willingness to use AI in both low- and high-stakes scenarios. We used a mixed-methods approach to collect both qualitative and quantitative data through interviews ($N = 12$) and a census-representative survey ($N = 302$) with end-users. The quantitative results of this study show that certification labels can be a promising way to communicate the outcome of audits to end-users, increasing both trust and willingness to use AI in low- and high-stake AI scenarios. Based on the qualitative findings, we further identified opportunities and limitations of certification labels, as well as inhibitors and facilitators for the effective design and implementation of certification labels. Our work provides the first empirical evidence that labels may be a promising constituent in the more extensive "trustworthiness ecosystem" for AI.

\section{Funding, declaration of conflicting interests and data availability}

This research was primarily funded by an independent research group, but additional funding (CHF 2,500.00) was granted by the Swiss Digital Initiative, an independent non-profit foundation, to obtain a representative sample. The entire research process, including the development of the research design, data analysis, interpretation of the results, and the writing of this paper, was conducted exclusively by independent researchers with no other affiliations with the Swiss Digital Initiative Foundation than those mentioned here. All data, corresponding R-scripts, and supplementary materials are available on OSF: \url{https://osf.io/gzp5k/}. 

\begin{acks}

Special thanks to Ariane Haller and the Swiss Digital Initiative for the permission to use their label for the purpose of our study, especially Nicolas Zahn, who was our contact person at the foundation. 

\end{acks}

\bibliographystyle{ACM-Reference-Format}
\bibliography{sample-base}

\appendix
\section{Appendix}
\label{appendix:a}

\subsection{High-stake Scenarios}

\subsubsection{Medical Diagnosis}

Consider the situation where you are searching for potential medical diagnoses.
Your insurance is using an AI system called MyHealth for evaluating medical symptoms. You will be required to fill out a form, uploading your medical history, and submit them along with personal information like age, gender, marital status and employment status to MyHealth. Once assessed, MyHealth will determine based on the provided information what your medical diagnosis is.

\subsubsection{Hiring Procedure}

Consider the situation where you are applying for a new job at a company. 
The company is using an AI system called MyJob for evaluating job applications. You will be required to fill out a form, uploading your CV, and submit them along with personal information like address, marital status, employment status and references to MyJob. Once assessed, MyJob will determine based on the provided information whether or not you will be invited for an interview.

\subsubsection{Loan Approval}

Consider the situation where you are applying for a loan at a bank. 
The bank is using an AI system called MyLoans for evaluating loan applications. You will be required to fill out a form, specifying the loan amount, and submit them along with personal information like marital status, employment status, annual income and financial history to MyLoans. Once assessed, MyLoans will determine based on the provided information whether your loan application is successful or not.

\subsection{Low-stake Scenarios}

\subsubsection{Music Preference}

Consider the situation where you want to explore new music.
You are using an AI system called MyMusik for evaluating your music preference. You will be required to accept terms and conditions of MyMusik which among other things include analyzing your search behavior and already liked songs. Once assessed, MyMusik will provide you with song recommendations.

\subsubsection{Route Planning}

Consider the situation where you want to get from one place to another place.
You are using an AI system called MyMap for evaluating your travelling route. You will be required to accept terms and conditions of MyMap which among other things include analyzing your motion data and already visited places. Once assessed, MyMap will provide you with a route recommendation.

\subsubsection{Price Comparision}

Consider the situation where you want to sell your car.
You are using an AI system called MyCar for evaluating a selling price. You will be required to accept terms and conditions of MyCar which among other things include analyzing your search history on the platform and already sold cars. Once assessed, MyCar will provide you with a selling price recommendation.

\end{document}